\documentclass{aa}
\usepackage{graphicx}
\newcommand{\PSR}{\object{PSR~J1740$-$5340}}
\newcommand{\NGC}{\object{NGC 6397}}
\newcommand{\U}{\ifmmode U_{336}\else$U_{336}$\fi} 
\newcommand{\B}{\ifmmode B_{439}\else$B_{439}$\fi}
\newcommand{\V}{\ifmmode V_{555}\else$V_{555}$\fi}
\newcommand{\R}{\ifmmode R_{675}\else$R_{675}$\fi}
\newcommand{\I}{\ifmmode I_{814}\else$I_{814}$\fi}
\newcommand{\Halph}{\ifmmode{\rm H}\alpha_{656}\else${\rm H}\alpha_{656}$\fi} 

\begin{document}
\title{The eclipsing millisecond pulsar PSR J1740$-$5340 and its red
straggler companion\thanks{Based on observations made with ESO
Telescopes at the Paranal Observatories under programme 267.D-5716 and
observations made with the NASA/ESA Hubble Space Telescope, obtained
from the ST-ECF data archive.}}

\author{J. A. Orosz\thanks{Present address: Astronomy Dept., San
Diego State Univ., 5500 Campanile Drive, San Diego, CA
92182-1221, USA; \email{orosz@sciences.sdsu.edu}}
\and    M. H. van Kerkwijk\thanks{Present address: Dept. of
Astronomy and Astrophysics, Univ.\ of Toronto, 60 St George
Street, Toronto, ON~~M5S~3H8, Canada; \email{mhvk@astro.utoronto.ca}}}

\offprints{J. Orosz}

\institute{Astronomical Institute, Utrecht University,
           P.O.\ Box 80\,000, 3508 TA Utrecht,  The Netherlands}

\date{Received ; accepted }

\abstract{We present a high-resolution echelle spectrum taken with the
Very Large Telescope and analyse archival {\em Hubble Space Telescope}
photometry of the recently identified companion of the eclipsing
millisecond radio pulsar \PSR\ {in the globular cluster \NGC}.
From the spectrum, we show that the companion is metal poor,
{as expected for a member of \NGC}.  Using synthetic
photometry, and assuming a true distance modulus of $12.13\pm0.15$ mag
and a colour excess of $E_{B-V}=0.179$, we derive a radius of
$1.60\pm0.17\,R_{\odot}$ and an effective temperature of
$5410\pm50{\rm\,K}$, implying a luminosity of $2.0\pm0.4\,L_\odot$
($1\sigma$ errors).  These properties make it similar to the so-called
`sub-subgiants' in the old open cluster \object{M\,67} and `red
stragglers' in \object{47~Tuc}, which have luminosities comparable to
those of turn-off stars, but cooler temperatures and larger radii.
The light curve of the companion is well described by ellipsoidal
variations, {and despite the incomplete ($\sim\!60\%$) phase
coverage, we are able to derive good constraints on a number of the
system parameters}.  In particular, for the inclination we find a
2-$\sigma$ lower limit of $i>48^{\circ}$ ($i>46^\circ$ at 3$\sigma$).
Assuming a pulsar mass of $1.2<M_{\rm MSP}<2.4\,M_{\odot}$, this
implies a companion mass in the range $0.14<M_{\rm
comp}<0.38\,M_{\odot}$.  Combined with the photometric constraint, we
find a best fit for $i\simeq50^\circ$ and $M_{\rm
comp}\simeq0.3\,M_\odot$.  We infer a Roche lobe filling factor by
radius of $\sim\!97$\%.

Surprisingly, we find no evidence whatsoever for irradiation of the
companion, despite the high inferred rotational energy loss of the
pulsar ($\dot{E} \simeq 1.4\times10^{35}{\rm\,erg\,s^{-1}}$).  We
discuss possible reasons, but find most lacking.  We hypothesise that
the system is a triple, and that the acceleration due to a third body
in a wide orbit around the binary led to an overestimate of the
intrinsic spin-down rate and hence the spin-down luminosity.  This can
be tested by further timing observations.  We also discuss two other
puzzles, viz., the system's location far outside the cluster core and
the companion's large radius and luminosity.  We suggest that the
system was formed in a binary-binary encounter in the core, due to
which the system acquired a substantial velocity, and the companion --
which must have been a somewhat evolved turn-off star -- lost much of
its envelope.  We suggest other `red stragglers' and `sub-subgiants'
might have formed by similarly drastic encounters.
}

\titlerunning{PSR J1740$-$5340 and its red straggler companion}
\maketitle


\section{Introduction}\label{intro}

In the course of a survey for pulsars in globular clusters, D'Amico et
al.\ (\cite{dalm+01}) discovered an unusual millisecond pulsar
(MSP), \PSR,
in the second-closest globular cluster, \NGC.  The pulsar has a
3.65\,ms spin period, and is in a circular ($e<10^{-4}$), 1.35-d orbit
around a companion with a minimum mass of $0.19\,M_\odot$ (assuming a
$1.4\,M_\odot$ neutron star).  In follow-up observations, D'Amico et
al.\ (\cite{dapm+01}) found that at a frequency of 1.4\,GHz, the
pulsar signal disappears for about 40\% of the orbit around superior
conjunction, and in addition shows irregular intensity and
arrival-time variations.  By selecting unaffected data sets, D'Amico
et al.\ were nevertheless able to derive a good timing solution,
including a precise position, which put the system at $0\farcm55$, or
11 core radii out of the core.

At the timing position, Ferraro et al.\ (\cite{fpdas01}) identified a
relatively bright ($V\approx17$), anomalously red star, which varied
at the orbital period.  The light curve appeared ellipsoidal, i.e.,
consistent with what would be expected for a tidally distorted star.
{This star had been discussed earlier by Taylor et al.\
(\cite{tgec01}), who did not know about the pulsar, but had noted that
the companion was a peculiar, variable star with a H$\alpha-R$ colour
indicating weak H$\alpha$ emission.  Taylor et al.\ also noted that
its proper motion was consistent with membership.}  Finally, Grindlay
et al.\ (\cite{ghe+01}) identified an X-ray counterpart, with a
luminosity of $\sim\!8\times10^{30}{\rm\,erg\,s^{-1}}$.

The system raises many interesting questions, such as:
\begin{enumerate}
\item How could the system have gotten so far from the core of the
globular cluster?  Given their high mass, pulsar systems should
quickly settle in the core, as this is indeed generally where they are
found.  Did the system get kicked out due to a close encounter?
\item How can one understand the strange position of the companion in
the colour-magnitude diagram?  The star is as luminous as a turn-off
star, but redder.  As pointed out by Edmonds et al.\ (\cite{egc+02}),
these properties are similar to what is seen for the `red stragglers'
in \object{47~Tuc} and `sub-subgiants' in the old
open cluster \object{M 67}.
\item What mechanism could cause a wind strong enough to eclipse the
pulsar emission?  In other eclipsing pulsars, it is thought the pulsar
irradiation drives a wind, but the light curve for \PSR\ shows no
indication for heating.  Could the wind be intrinsic?
\end{enumerate}

Fortunately, the companion is bright, and hence detailed studies are
possible.  This opens the possibility not only of addressing the above
questions, but also of measuring the system parameters accurately.
The latter would be particularly interesting since no accurate masses
exist for MSPs.  As MSPs are expected to be more massive due to
accretion, this might be interesting also from the point of view of
constraining the equation of state of ultra-dense matter {(for
a review, see Lattimer \& Prakash \cite{lp01})}: neutron stars with
masses above {$\sim\!1.6\,M_\odot$} cannot exist for so-called
soft equations of state, in which matter at high densities is
relatively compressible (e.g., due to meson condensation or a
transition between the hadron and quark-gluon phases).  Obviously, if
one were to have solid evidence that massive neutron stars exist,
these equations of state would be falsified.

Given the above, a thorough observing programme seems justified.  In
this paper, we report on the analysis of archive photometry and
present the results of a pilot programme designed to test the
feasibility of high resolution spectroscopic observations.

\section{Observations}

\begin{figure}
\centering
\includegraphics[width=8.8cm]{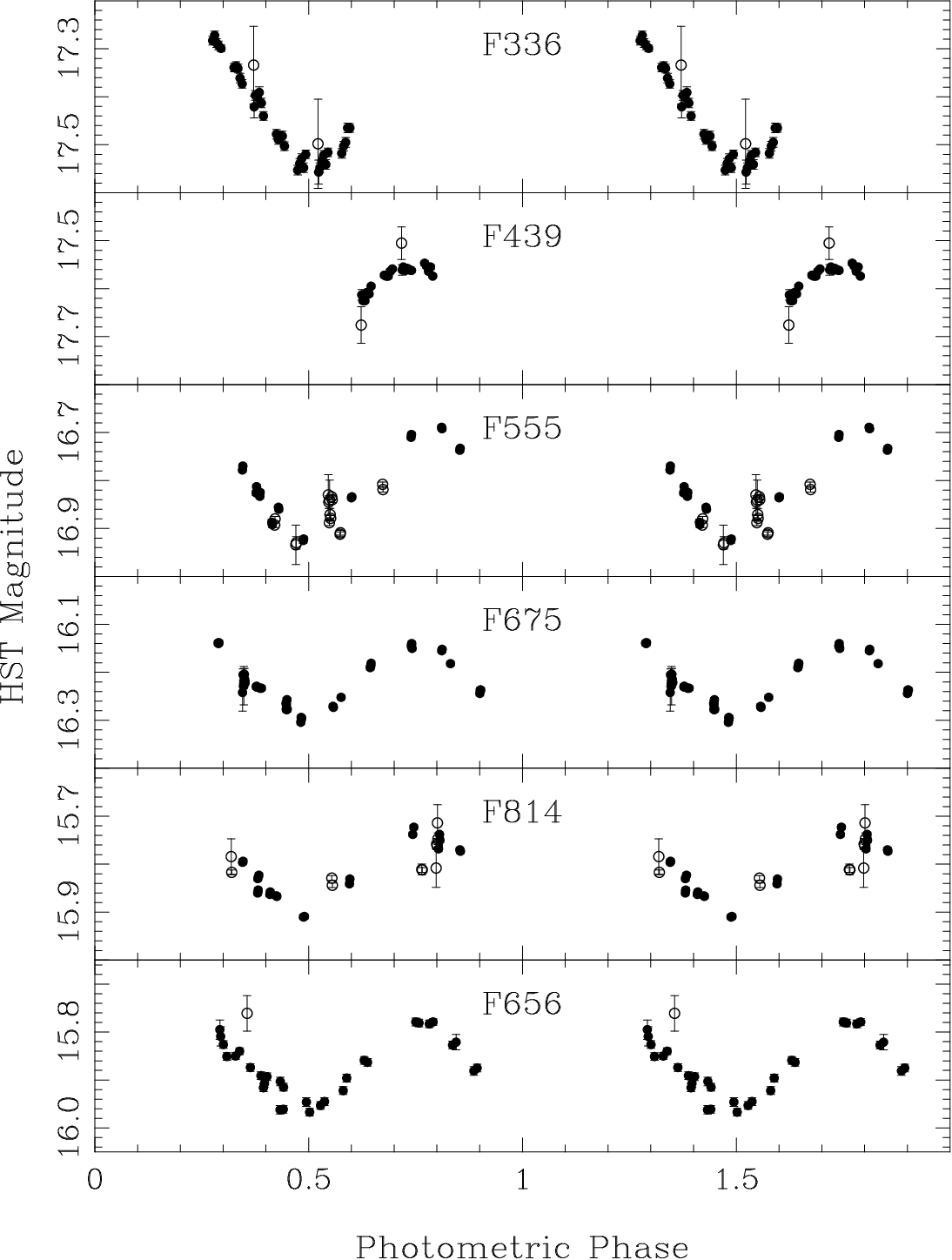}
\caption{The {\em HST} data as a function of photometric phase.  Phase
zero is the time of the inferior conjunction of the companion star.
The points plotted with the open circles are excluded from analysis.}
\label{showlc}
\end{figure}

\subsection{Photometry}

\begin{figure}
\includegraphics[width=8.8cm]{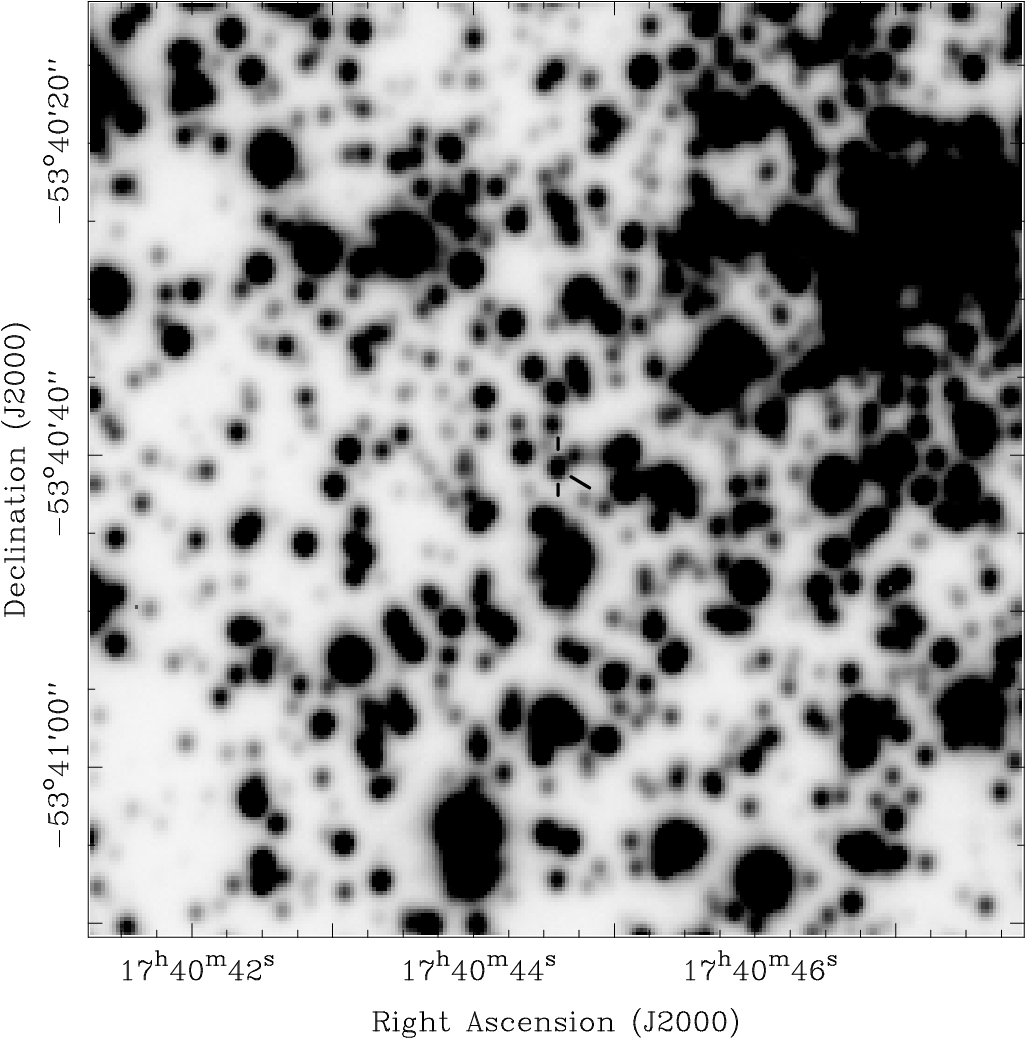}
\caption{Finding chart for \PSR.  This image is a $1\times 1$
arcminute subsection of a $V$-band image obtained May 14, 1999 with
the Wide Field Imager on the ESO 2.2m telescope at La Silla.}
\label{figfind}
\end{figure}

The photometric data we used were taken from the public {\em HST}
archive.  The observations were carried out on 1996 March in the
F336W, F439W, F555W, and F814W filters (hereafter \U, \B, \V, and \I)
and 1999 April using the \V, F675W (hereafter \R), \I, and F656N
(hereafter \Halph) filters.  Light curves of the companion were
constructed using the HSTphot package (version 1.1; Dolphin
\cite{dol00}).

In Fig.~\ref{showlc}, we show the folded light curves in all bands.
As discussed by Ferraro et al.\ (\cite{fpdas01}), {the overall
shape of the lightcureve is reminiscent of ellipsoidal variations, in
which the maxima are at the orbital quadratures, when the tidally
distorted companion is seen from the side}.  We find that the curves
in \U, \B, \R, and \Halph\ curves are reasonably smooth and show the
ellipsoidal modulations quite nicely, but that those in \V\ and \I\
have some deviant points.  We believe the latter are due to the very
close proximity of a bad column on the CCD.  Indeed, the \R\ and
\Halph\ band data from MJD 51273 show smooth variations whereas the
\V\ and \I\ data from that same day both show deviant points.  Also,
the \V\ and \I\ light curves are the only ones with data from both
1996 and 1999.  Removing the 1996 data improves the \I\ band light
curve somewhat but does not really improve the \V\ light curve.

The photometric data that we will use in our analysis below are shown
as the solid points in Fig.\ \ref{showlc}.  The excluded \V\ data are
those from 1996 plus eight points from MJD $51\,272.807-51\,272.819$.
The excluded \I\ data are those from 1996 plus five points from MJD
$51\,271.794-51\,271.800$ and two points from near MJD $51\,272.818$.
Also, two points in \U, two points in \B, and one point in \Halph\
were excluded owing to the large uncertainties.  

{As is clear from Fig.~\ref{showlc}, the phase coverage of the
{\em HST} light curves is not complete, and clearly additional
photometry would be desirable}.  Since the field of the pulsar is
relatively crowded, we present in Fig.~\ref{figfind} a finding chart
as an aid for ground-based observers.  This image was obtained May 14,
1999 with the Wide Field Imager on the 2.2m telescope at ESO, La
Silla.

\subsection{Spectroscopy}

A 2880-second UVES spectrum in 0.8 arcsecond seeing was obtained for
us in Director's Discretionary Time on 2001 September 20, starting at
23:56 UT.  UVES resides at the Nasmyth focus of Kueyen, the second of
the four 8.2-m Very Large Telescopes at Paranal (Dekker et al.\
\cite{ddok+00}; D'Odorico et al.\ \cite{docd+00}).  UVES is a
double-arm instrument, with a $2048\times4096$ thinned,
antireflection-coated CCD (EEV CCD-44, $15\,\mu$m pixels) in the blue
arm and a mosaic of two $2048\times4096$ CCDs (EEV CCD-44 and MIT/LL
CCD-20, both with $15\,\mu$m pixels) in the red arm.  In this paper we
will only make use of the data from the red arm.  The standard ``DIC1
346+580'' configuration was used, which gives a wavelength coverage of
4765-6830~\AA\ in the red arm.  With a 1\arcsec\ wide slit, the
resolving power is about $40\,000$.  The CCDs were read out in the
$2\times2$ pixel binned mode.  We used IRAF to calibrate the data and
extract the spectra.  The typical signal-to-noise ratio is about 20
per resolution element.

\section{Analysis}

\subsection{Radius and temperature}

\begin{figure}
\centering
\includegraphics[width=8.8cm]{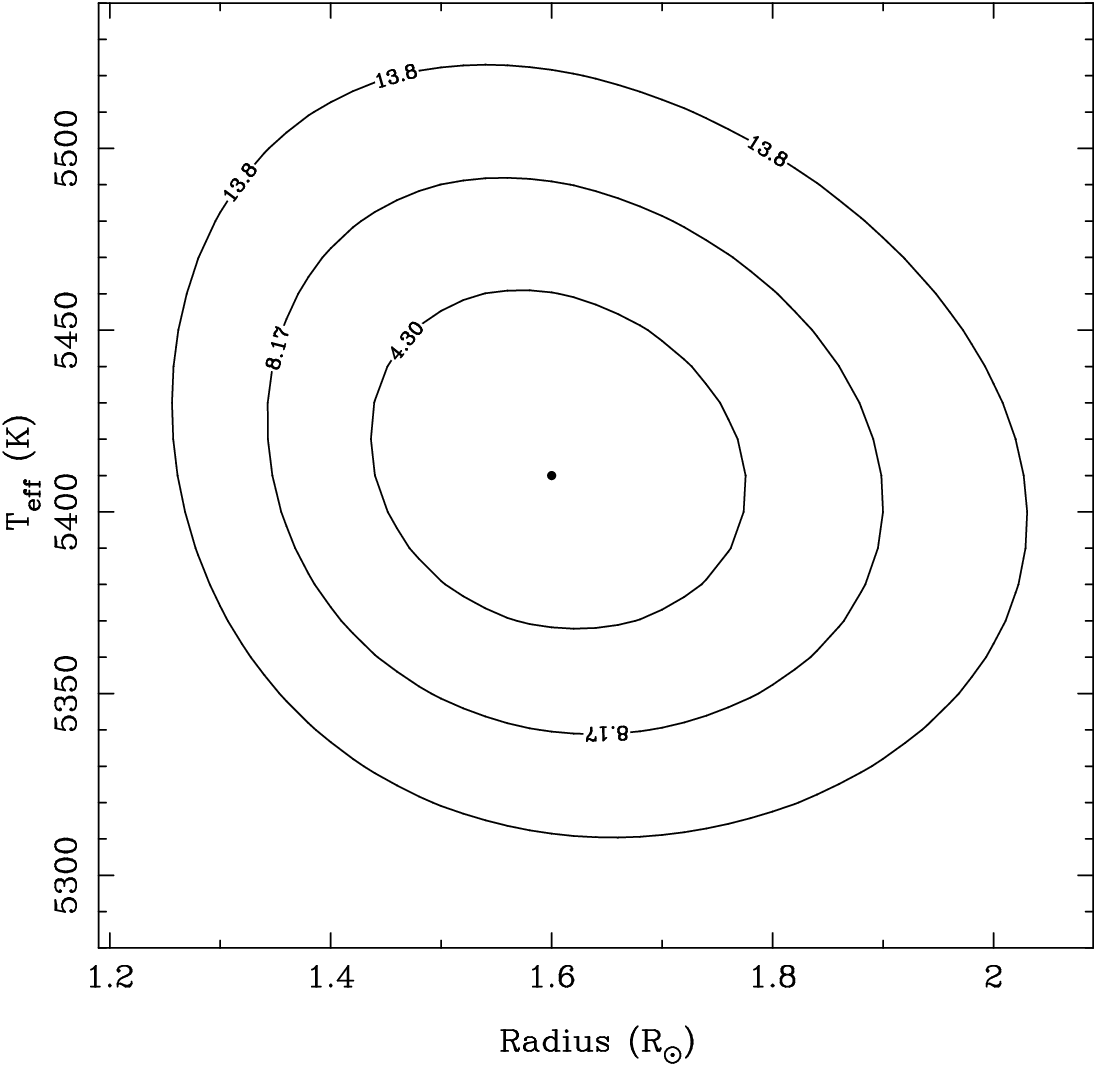}
\caption{The radius and temperature of the companion star derived by
comparing the {\em HST} magnitudes and colours with synthetic
photometry from the {\sc NextGen} model atmospheres (Hauschildt et
al.\ \cite{hba97}; F.\ Allard 2001, priv.\ comm.).  The contours show
the 1, 2, and $3\sigma$ confidence limits for two parameters of
interest ($\Delta\chi^2=4.30$, 8.17, and 13.8, respectively).  We find
$R_{\rm comp} = 1.60\pm0.17\,R_\odot$ and $T_{\rm eff} =
5410\pm50{\rm\,K}$, assuming a reddening of $E_{B-V}=0.179$
(Anthony-Twarog \& Twarog \cite{att00}) and a true distance modulus of
$12.13\pm0.15$ mag (Reid \& Gizis \cite{rg98}; {see text for a
discussion on the distance}).}
\label{plotrad}
\end{figure}

We can use the fact that \PSR\ is a member of the globular cluster
\NGC\ (see also Section \ref{spectcomp}) to measure the radius and
effective temperature of the companion.  For this exercise we use
synthetic photometry for the {\em HST} filters derived from the {\sc
NextGen} models (Hauschildt et al.\ \cite{hba97}; F.\ Allard 2001,
priv.\ comm.) with a metallicity of ${\rm [Fe/H]}=-2$, appropriate for \NGC\
(e.g., Castilho et al.\ \cite{cpa+00}; Th\'evenin et al.\
\cite{tcdfp+01}).  The synthetic photometry allows us to compute the
expected absolute magnitude of a star in a given {\em HST} bandpass
and various colours as a function of its effective temperature $T_{\rm
eff}$, surface gravity $\log g$, and radius $R_{\rm comp}$.

To compare the synthetic magnitudes with the observations, we need to
determine the `mean' apparent magnitudes.  For this purpose, we used
the ELC code (Orosz \& Hauschildt \cite{oh00}), as follows.  First, we
fit an ellipsoidal model (see Sect.~\ref{sec:model}) to the light
curve in the \R\ filter (which has the most complete phase coverage).
We then computed the flux of a spherical star with the same effective
temperature and effective radius and applied the same scaling as the
fitted ellipsoidal model.  The spherical star would have an \R\
magnitude of 16.25.  

{To convert the above apparent magnitude to an absolute one,
we need to know the reddening and the distance.  For the reddening, we
adopt $E_{B-V}=0.179\pm0.003$, as determined by Anthony-Twarog \&
Twarog (\cite{att00}) from $uvby$H$\beta$ photometry, and use the
interpolated extinction curve for the WFPC2 filters appropriate for a
K5 spectral type (Holtzman et al.\ \cite{hbc+95}).  For the distance,
we use a distance modulus $(m-M)_0=12.13\pm0.15$, as inferred by Reid
\& Gizis (\cite{rg98}) from main-sequence fitting to local metal-poor
M subdwarfs with well-determined parallaxes.  

We note that the uncertainty in the distance dominates, not only
because of the large random error, but also because of possible
systematic biases.  Indeed, from different methods rather different
values are found.  In an earlier analysis based on subdwarf G and F
stars, Reid (\cite{rei98}) found a distance modulus of $12.24\pm0.10$,
i.e., a larger distance.  In contrast, Harris (\cite{har96}) finds a
smaller distance -- 11.80 with an estimated uncertainy of
0.1--0.2\,mag -- based on the apparent magnitude of the horizontal
branch.  Finally, Gratton et al.\ (\cite{gcb+02}) present preliminary
results from a more advanced main-sequence fit, based on precise
matching of spectral types with high-resolution spectroscopy, and
$E_{B-V}=0.183\pm0.005$, ${\rm [Fe/H]}=-2.03\pm0.04$, and
$(m-M)_0=12.01\pm0.06$.  For our analysis, we use the Reid \& Gizis
(\cite{rg98}) distance, since this seems the most reliable of the
published results, and is the one used by most authors so far.
However, where appropriate, we will mention explicitly the
implications of assuming the shorter Gratton et al.\ (\cite{gcb+02})
distance.}

With our adopted reddening and distance modulus, we find an absolute
magnitude in the F675W bandpass of $M_{675} =
(16.25-0.44-12.13)\pm0.15 = 3.64\pm0.15$, where the error in the
absolute magnitude was taken to be the error in the distance modulus.
A similar procedure was used to find the colours; we found
$\B-\V=0.867$, $\V-\R=0.597$, and $\R-\I=0.406$.

For a given combination of $T_{\rm eff}$, $\log g$, and $R_{\rm
comp}$, we can form a $\chi^2$:
\begin{eqnarray}
\chi^2 &=& {[(B-V)_{\rm syn}-0.867]^2\over \sigma_{\rm colour}^2} 
       + {[(V-R)_{\rm syn}-0.597]^2\over \sigma_{\rm colour}^2}\nonumber\\
       &+& {[(R-I)_{\rm syn}-0.406]^2\over \sigma_{\rm colour}^2} 
       + {[M_{\rm 675,syn}-3.64]^2\over 0.15^2}.
\end{eqnarray}
Here, $\sigma_{\rm colour}$ is the uncertainty in the colours; we
adopt $\sigma_{\rm colour}=0.0144$, which gives $\chi^2_{\nu}=1$ at
the minimum.  We found that the magnitude and colours hardly depend on
gravity, so we fixed $\log g$ at its nominally best value of 3.7.
This leaves the effective temperature $T_{\rm eff}$ and radius $R_{\rm
comp}$ as the two free parameters.  Figure \ref{plotrad} shows the
contours of $\chi^2$ in the $R_{\rm comp}-T_{\rm eff}$ plane.  We find
$R_{\rm comp} = 1.60\pm0.17\,R_{\odot}$ and $T_{\rm eff} =
5410\pm50$~K.  These values are consistent with what Ferraro et al.\
(\cite{fpdas01}) found by independent means.  {Assuming the
Gratton et al.\ (\cite{gcb+02}) distance modulus, the inferred radius
and temperature would be $1.52\pm0.08\,R_\odot$
and $5420\pm 40\,K$, respectively.}
   
\subsection{Radial and rotational velocities, and metallicity}
\label{spectcomp}

\begin{figure}
\includegraphics[width=8.8cm]{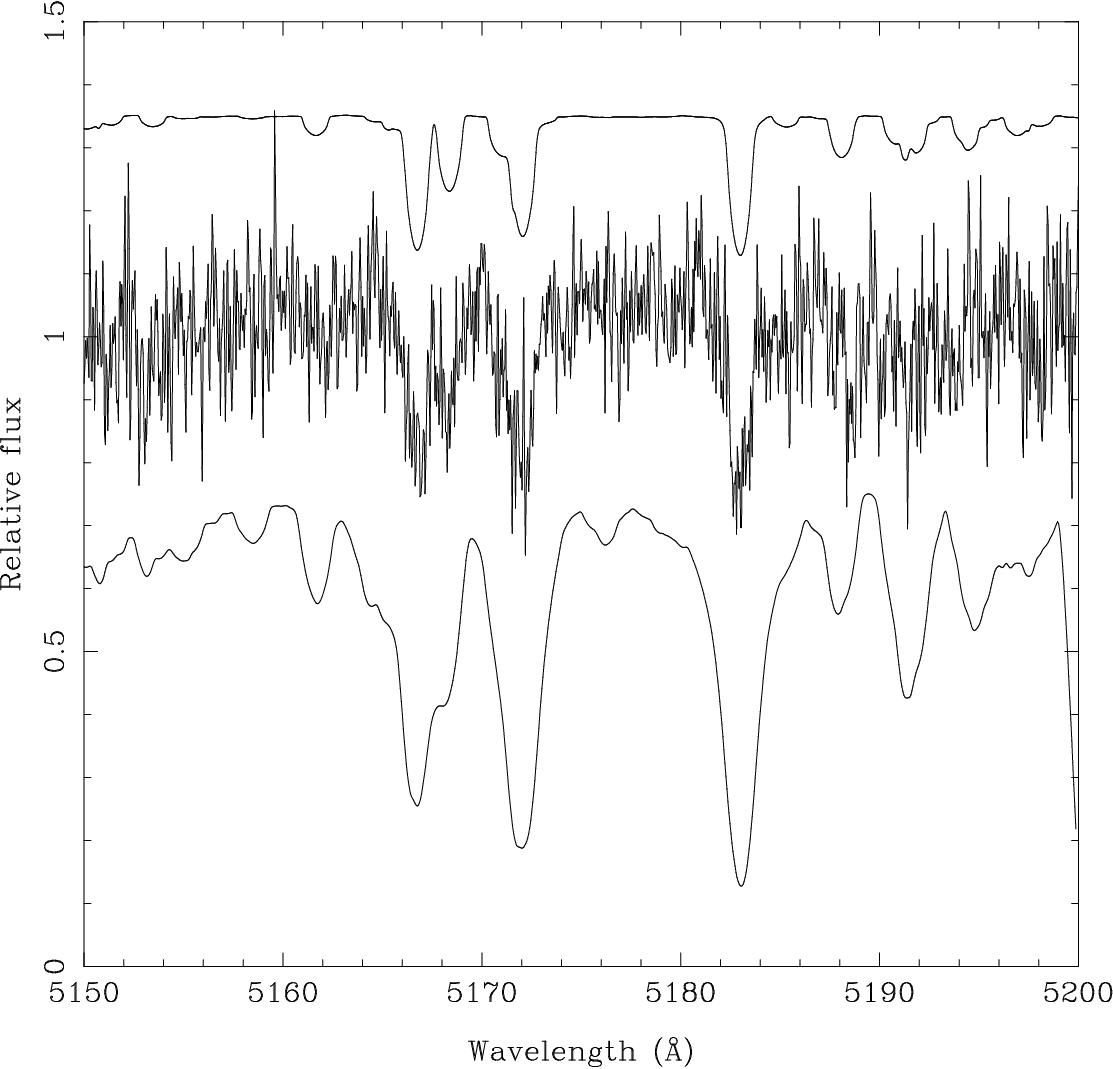}
\caption{UVES spectra of the companion of \PSR\ (middle), compared
with spectra broadened to $v_{\rm rot}\sin i=52{\rm\,km\,s^{-1}}$ of
the metal-poor star HD 122196 (top; $T_{\rm eff}=5850\pm 100$~K, $\log
g=3.5$, [Fe/H]$=-1.93\pm 0.10$; Ryan \& Deliyannis \cite{rd98}), and
of the solar-metallicity star HR 996 (bottom; $T_{\rm eff}=5667$~K,
$\log g=4.29$, [Fe/H]$=-0.01$; Pasquini et al.\ \cite{plp94}, taken
from the atlas of Montes \& Martin \cite{mm98}).}
\label{plotspec}
\end{figure}

In Fig.~\ref{plotspec}, we show the spectrum of \PSR\ near the Mg b
lines.  The lines are clearly broadened, as expected if the star is
co-rotating with the 1.35-d orbit.

We measured the radial and rotational velocities of the companion star
using a template, for which we chose the metal poor star HD 122196
{($V=8.73$, $E_{B-V}=0.01$)}.
This star has an effective temperature and gravity ($T_{\rm
eff}=5850\pm 100$~K, $\log g=3.5$; Ryan \& Deliyannis \cite{rd98})
quite similar to what we found for the companion star of \PSR, as well
as a metallicity (${\rm[Fe/H]}=-1.93\pm0.10$; ibid.) similar to that
of \NGC\ members.  Furthermore, this star was observed with UVES using
the same instrumental configuration as what was used for \PSR,
although at a slightly higher resolving power ($R=50\,000$).

To measure the radial velocity by cross correlation, we have to
iterate a bit, since we have to rotationally broaden the template to
match the rotational velocity of the companion.  Given a radius of
$R_{\rm comp}=1.60\,R_{\odot}$ and assuming synchronous rotation, one
expects $v_{\rm rot}\sin i=42.3{\rm\,km\,s^{-1}}$ for an inclination
of $45^{\circ}$ and $59.8{\rm\,km\,s^{-1}}$ for an inclination of
$90^{\circ}$.  We initially broadened the spectrum of HD 122196
{(which has negligible intrinsic rotational broadening)} by
$49{\rm\,km\,s^{-1}}$ and measured the radial velocities of each \PSR\
echelle order with the IRAF task fxcor, an implementation of the Tonry
\& Davis (\cite{td79}) cross correlation technique.  Good cross
correlation peaks were found for 14 echelle orders, and the weighted
average of the (heliocentric) velocities from each order is
$135\pm1{\rm\,km\,s^{-1}}$ (assuming HD 122196 has a heliocentric
velocity of $-23{\rm\,km\,s^{-1}}$).  

To find the rotational velocity of the companion, the spectrum of HD
122196 was Doppler shifted to remove the relative velocity difference
between it and the \PSR\ spectrum, and the $\chi^2$-based matching
technique of Marsh, Robinson, \& Wood (\cite{mrw94}) was used to
measure the projected rotational velocity.  Using the spectral region
near the Mg b lines, which contains the strongest lines, we find
$v_{\rm rot}\sin i = 52\pm4{\rm\,km\,s^{-1}}$.  This value should be
treated with caution owing to the relatively low signal-to-noise and
the fact that the template spectrum may not be a perfect match (e.g.,
owing to differences in the Mg and/or Fe abundances).  We re-measured
the radial-velocity measurement using this new value of $v_{\rm
rot}\sin i$, but found that it did not change the result.

In Fig.~\ref{plotspec}, we compare the spectrum of \PSR\ and the
spectrum of HD 122196, broadened using $V_{\rm rot}\sin i=52$ km
s$^{-1}$.  To the extent one can judge, given the rather noisy
spectrum, the match is reasonable.  For comparison, we also show a
high resolution spectrum of HR 996, a star with a similar temperature
as \PSR, but with solar metallicity.  The Mg b lines in HR 996 are
clearly much stronger than they are in \PSR.  Given this and the
reasonable match between \PSR\ and the metal poor star HD~122196, we
conclude that the companion in \PSR\ is metal poor, {providing
further confirmation} that it is a member of the globular cluster
\NGC\ and not a foreground object in the Galactic disc.

Assuming the MSP is bound to the cluster, our single radial velocity
measurement allows us to make a {\em rough} estimate of the radial
velocity amplitude $K_{\rm comp}$.  This is because the cluster
membership constrains the systemic velocity~$\gamma$, and thus we can
calculate $K_{\rm comp}$ from $K_{\rm comp}=(v_{\rm
obs}-\gamma)/\sin(2\pi\phi)$, where $\phi=0.368$ is the phase relative
to inferior conjunction of the companion.  Given the radial velocity
of \NGC\ of $18.9\pm0.1{\rm\,km\,s^{-1}}$ (Harris \cite{har96}) and
its escape velocity of $\sim\!19{\rm\,km\,s^{-1}}$ (Webbink
\cite{web84}), we take $0<\gamma<38{\rm\,km\,s^{-1}}$, and infer
$132<K_{\rm comp}<183{\rm\,km\,s^{-1}}$.  The pulsar timing yields
$K_{\rm MSP}=26.612{\rm\,km\,s^{-1}}$ (D'Amico et al.\
\cite{dapm+01}), and hence the mass ratio $Q\equiv M_{\rm MSP}/M_{\rm
comp}$ should be in the range $5<Q<7$.  Obviously, one should not rely
on a single radial velocity point to measure the amplitude of the
companion star's velocity curve.  Nevertheless, this exercise at least
gives a rough idea of what the mass ratio might be.  As we will see
below, the values we find are consistent with what we infer from the
other data on the system.

\begin{figure*}
\centering
\includegraphics[angle=-90,width=18cm]{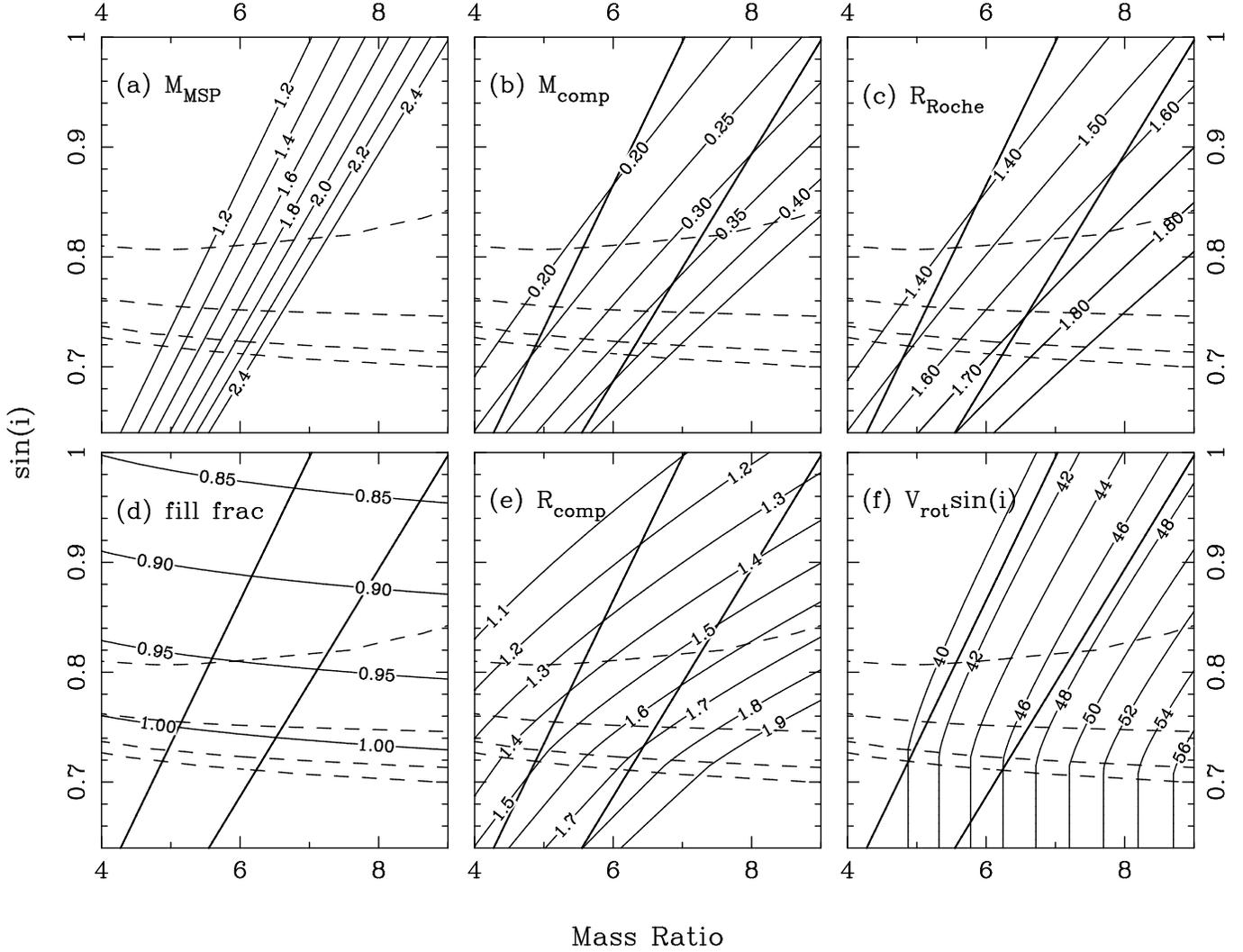}
\caption{Constraints on the system parameters in the mass-ratio,
inclination plane.  In all panels, the dashed lines indicate, from top
to bottom, the $1\sigma$, $2\sigma$, $3\sigma$, and 99.99\% confidence
limits inferred from our ellipsoidal fits to the six {\em HST} light
curves.  The fits set a stringent lower limit to the inclination, but
hardly constrain the mass ratio.  
{\bf a)} Contours of the MSP mass (in solar masses).  For reasonable
neutron star masses ($1.2\,M_{\odot}-2.4\,M_{\odot}$) a relatively
small part of the plane is filled.  In all other panels, the thick
diagonal lines denote the borders of the region where $1.2\le M_{\rm
MSP}\le 2.4\,M_{\odot}$. 
{\bf b)} Contours of the companion star mass (in solar masses).
{\bf c)} Contours of constant Roche-lobe radius of the companion (in
solar units).  
{\bf d)} Contours of constant best-fitting filling fraction of the
companion star (by radius, i.e., $R_{\rm comp}/R_{\rm Roche}$).
{\bf e)} Contours of the constant companion star radius corresponding
to the best-fitting filling factor (in solar units).
{\bf f)} Contours of constant projected rotational velocity
corresponding to the best-fitting companion radius (in ${\rm
km\,s^{-1}}$).}
\label{plotsincont}
\end{figure*}

\subsection{Ellipsoidal modelling}\label{sec:model}

{The {\em HST} lightcurves of \PSR\ are, as mentioned above,
reminiscent of ellipsoidal modulation, with maxima at the quadratures,
and minima at the conjunctions.  Below, we will use models for a
tidally distorted star to reproduce the lightcurve and to estimate
some of the system parameters.  Before doing so, we should caution
that the phase coverage is incomplete, and that, in particular,
superior conjunction of the pulsar is missing.  Thus, even though we
obtain good fits to the data at hand, whether or not our models are
appropriate, remains to be fully verified.}

For our modeling of the {\em HST} light curves, we used the ELC code
(Orosz \& Hauschildt \cite{oh00}).  We have three unknown geometrical
parameters: the mass ratio $Q$, the inclination $i$, and the Roche
lobe filling factor by radius $f$.  Usually, one also has the orbital
separation $a$ as an unknown.  For \PSR, however, the projected
semi-major axis of the MSPs orbit is known extremely accurately
($x\equiv a_{\rm MSP}\sin i/c = 1.65284\pm0.00007{\rm\,lt-s}$; the
$1\sigma$ error corresponds to about 7\,km; D'Amico et al.\
\cite{dapm+01}).  As a result, we can write the orbital separation as
a function of the inclination and the mass ratio, $a = c(Q+1)x/\sin
i$.

We also require parameters related to the companion.  These, however,
can be fixed at reasonable values.  We assume the companion star is
rotating synchronously with the orbit, fix the mean temperature at the
value determined above, and fix its gravity darkening exponent at
$\beta=0.1$ (Claret \cite{cla00}).  We used specific intensities for
models with $[{\rm Fe/H}]=-2$ from
Kurucz\footnote{http://www.cfaku5.harvard.edu} (\cite{kur79}),
integrated with the WFPC2 filter response curves.  The use of the
tabulated specific intensities removes the need for a parameterised
limb darkening law.

Finally, we need to consider irradiation by the pulsar.  As we will
discuss at length in Sect.~\ref{noheating}, this should have had a
dominant effect on the light curve, but no effect whatsoever is seen.
For the remainder of this section, therefore, we will ignore it.

In any fitting procedure, one must assign weights to the observations.
We do this by fitting an ellipsoidal model to each bandpass
separately, and scaling the error bars of the data points in that
bandpass so that the reduced $\chi^2$ of the fit equals 1 at the
minimum.  The scaled errors are typically between about 0.015 and
0.022 magnitudes, which is on the order of the claimed accuracy of
HSTphot (Dolphin \cite{dol00}), and thus provides some evidence that
our model is appropriate.  Next, using these scaled errors, we fit all
of the bandpasses simultaneously.  With this procedure, somewhat
reduced weight is given to bands in which the light curves are
relatively ratty, such as \V\ and~\I.  {For our best fits, we
find $\chi^2\simeq225$ for 172 data points in the 6 filters, i.e., a
reduced $\chi^2$ which is still reasonably
close to unity.  This shows that best
fits to the lightcurves in the different bands are mutually
consistent.}

For our fits, we found it convenient to define a grid of points in the
mass ratio-inclination plane and to make the grid spacing in the
inclination axis equal in $\sin i$ rather than $i$.  At each point in
this plane, the values of the mass ratio and the sine of the
inclination are fixed according to the location in the plane, leaving
the Roche lobe filling factor $f$ as the only free fitting parameter.
In Fig.~\ref{plotsincont}, we show the results of the ellipsoidal
fitting as the dashed contours.  We find that good fits are possible
over a large range of inclinations: the formal $2\sigma$ range
includes all $i>49.5^{\circ}$ ($\sin i > 0.76$).  The mass ratio $Q$
is not constrained.

In Fig.~\ref{plotsincont}d, solid contours show the variation of the
best-fitting filling factor in the $Q-\sin i$ plane (here, the filling
fraction is defined as $R_{\rm comp}/R_{\rm Roche}$).  From these, the
reason for the large allowed range in inclination becomes clear: the
amplitude of the ellipsoidal variations can be matched by stars viewed
almost edge-on that underfill the Roche lobe significantly and hence
are less strongly distorted, as well as by stars viewed more inclined
that almost fill the Roche lobe and thus are maximally distorted.

\begin{figure}
\includegraphics[width=8.8cm]{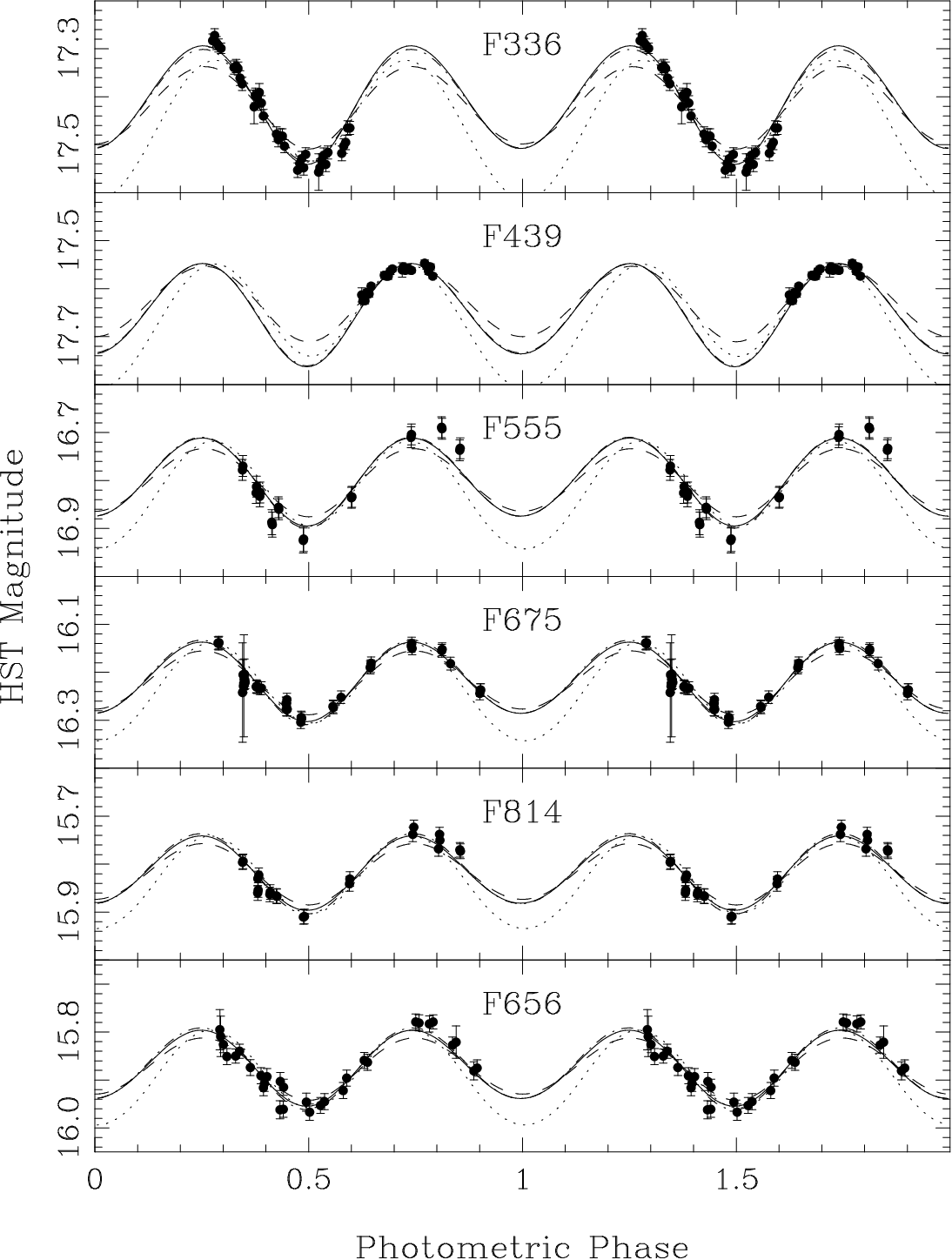}
\caption{Ellipsoidal models 
{with no X-ray heating} overdrawn on the {\em HST} light curves.
The solid lines are for a model with $i=60\fdg5$, $Q=6$, and a filling
fraction by radius of 91.1\% for the companion star.  The fit has
$\chi^2=225$ and is near the minimum.  The dash-dotted lines denote a
model with a Roche lobe filling companion, $i=46.7^{\circ}$, and
$Q=6$.  The fit has $\chi^2=236$ and is at the formal $2\sigma$ limit.
The dashed lines denote a model with a Roche lobe filling companion,
$i=40^{\circ}$, and $Q=6$.  The fit has $\chi^2=411$; it is clearly
excluded by the data.
{Finally, the dotted line denotes a model with a Roche lobe
filling companion, $i=52\fdg3$, $Q=6$, and an albedo of 10/
fit has  $\chi^2=411$, and is also clearly excluded by the data.}}
\label{showfitlc}
\end{figure}

The similarity in the fits can be see in Fig.~\ref{showfitlc}, which
shows some representative ellipsoidal fits overdrawn on the folded
{\em HST} light curves.  The first model is very near the formal
$\chi^2$ minimum, with $i=60\fdg5$, $Q=6$, and a filling fraction by
radius of 91.1\% for the companion star.  The second model is also for
$Q=6$, but for a Roche-lobe filling companion.  In this case,
$i=46\fdg7$.  Although the $\chi^2$ values are somewhat different
($\chi^2=225$ and $\chi^2=236$, respectively), the quality of the fits
looks very similar to the eye.

Turning back to Fig.~\ref{plotsincont}d, we see that because a star
cannot become larger than the Roche lobe, the fits get worse very
quickly for inclinations less than $49^{\circ}$: the $3\sigma$ contour
is at $i\approx 46^{\circ}$ ($\sin i\approx 0.72$) near $Q=6$ and the
99.99\% confidence limit is at $i\approx 45^{\circ}$ 
($\sin i\approx 0.71$).  At $i\approx 40^{\circ}$ ($\sin i\approx
0.64$; the bottom of all panels) the ellipsoidal fits are hopelessly
bad, as can be seen in Fig.~\ref{showfitlc}: the amplitudes are much
too small.  Thus, we regard $i>40^{\circ}$ as a very firm lower limit.

In the other panels of Fig.~\ref{plotsincont}, contours are drawn for
various other quantities.  The top row shows the MSP mass
(Fig.~\ref{plotsincont}a), the companion star
mass~(\ref{plotsincont}b), and the Roche lobe radius of the
companion~(\ref{plotsincont}c); these quantities can be calculated
from $Q$ and $\sin i$ directly, and the contours are thus independent
of the fit.  If we use the limit on the inclination, and restrict
ourselves to a reasonable range for the mass of the neutron star
($1.2\,M_{\odot}<M_{\rm MSP}<2.4\,M_{\odot}$), then we see from
panel~(a) that we are left with only a relatively small part of the
parameter space: the mass ratio is in the range $5\la Q\la 9$ for
$i\ge 45^{\circ}$.  Given this region where the neutron star mass is
reasonable, we see in Fig.~\ref{plotsincont}b that the mass of the
companion star is in the range $0.14\,M_{\odot}<M_{\rm
comp}<0.38\,M_{\odot}$.  From Fig.~\ref{plotsincont}c, the Roche
radius of the companion is smaller than the radius derived from {\em
HST} photometry for inclinations near $90^{\circ}$ unless the mass
ratio $Q$ is very large.

In the bottom row, quantities are shown that do depend on the 
light curve fit.
Fig.~\ref{plotsincont}e shows the best-fitting radius for the
companion star; this is simply the best-fitting filling fraction times
the Roche lobe radius.  We see that for high inclinations ($\sin i\ge
0.9$ or $i> 64^{\circ}$) the fitted radius of the companion star is
much smaller than the radius derived from the {\em HST} photometry.
On the other hand, the fitted radius is comparable to the radius
derived from the {\em HST} photometry for $i\approx50^{\circ}$ ($\sin
i\approx 0.77$) and $Q\approx6$.  This is also consistent with the
rough estimate of $Q$ from our single radial velocity. 

Finally, Fig.~\ref{plotsincont}f shows contours of the best-fitting
projected rotational velocity of the companion star (i.e.,
proportional to best-fitting radius times $\sin i$).  The contours bend
where the star starts to fill its Roche lobe.  We note that our
present value of $v_{\rm rot}\sin i=52{\rm\,km\,s^{-1}}$ is well
outside the range of parameter space where the mass of the MSP is
reasonable.  However, the statistical error on $v_{\rm rot}\sin i$ is
large (e.g., a $2\sigma$ deviation down to $V_{\rm rot}\sin
i=44{\rm\,km\,s^{-1}}$ would give a reasonable mass for the MSP) and,
as mentioned, there may be systematic errors as well.  The
uncertainties associated with the ellipsoidal modelling are smaller:
the 1-$\sigma$ uncertainty in the actual predicted rotational
velocity of the contours in Fig.~\ref{plotsincont}f is about
$0.1{\rm\,km\,s^{-1}}$. 

\section{The puzzling lack of heating}
\label{noheating}

We found that the light curve displayed by the companion is described
well by ellipsoidal modulation.  This is remarkable, as we would have
expected to see strong heating: the pulsar's rotational energy loss,
$L_{\rm sd} = I\Omega\dot\Omega =
1.4\times10^{35}I_{45}{\rm\,erg\,s^{-1}}$ (D'Amico et al.\
\cite{dapm+01}) is roughly 19 times the bolometric luminosity of the
companion.  We calculated models in which we included the effect of
heating, assuming isotropic emission.  If we assume that the companion
has an albedo or `reprocessing efficiency' with the canonical value of
0.5, then the expected \V\ light curve would be nearly sinusoidal with
an amplitude of 0.4 magnitudes and would have a {\em maximum} at phase
0.5, rather than the {\em minimum} that is observed.  {Even an albedo
of only 10\% leads to a horrible fit, as can be seen in
Fig.~\ref{showfitlc}}.  Indeed, if we leave the albedo as a free
parameter, we find that it must be 1\% or less in order to match the
observed light curves; the best-fitting value is zero.

The limit of 1\% is in striking contrast to what is seen {in
all pulsar systems in which irradiation could be observed.}  In
particular, for the so-called `black-widow' pulsars, which show light
curves dominated by heating, the inferred reprocessing efficiencies
are $\sim\!10$\% for \object{PSR~B1957+20} (Callanan et al.\
\cite{cvpr95}) and $\sim\!40$\% for \object{PSR~J2051-0827} (Stappers
et al.\ \cite{svkbk01}).  A high efficiency, of $\sim\!60\%$, has also
been inferred for the MSP \object{47~Tuc~U} and its low-mass white
dwarf companion (Edmonds et al.\ \cite{egh+01}), and strong heating is
observed for the pulsar, main--sequence star binary \object{47~Tuc~W}
(Edmonds et al.\ \cite{egc+02}).

What could be the cause of this apparent low efficiency?  The
following seem possible: (i) the pulsar's luminosity is lower than
inferred from its spin period and period derivative; (ii) the pulsar
radiation is emitted non-isotropically, largely missing the companion;
(iii) the pulsar radiation is absorbed or deflected on the way to the
companion; (iv) the radiation is absorbed sufficiently deep that it
can be transported, e.g., by winds as in Jupiter, and re-emitted
isotropically.  

On empirical grounds, we believe we can exclude possibilities (iii)
and (iv): no evidence is seen for either in the
{above-mentioned systems, which include companions ranging
from brown dwarfs to main-sequence stars and white dwarfs}.
Also anisotropy seems excluded: while pulsars winds are far from
isotropic, as is clear from the Crab and its nebula (e.g., Weisskopf
et al.\ \cite{wei++00}), it is hard to see how a companion could fail
to be irradiated over its whole orbit.  Indeed, the smooth light
curves of the black widow pulsars and symmetric bow shocks around the
MSP binaries \object{PSR~B1957+20} (Kulkarni \& Hester
\cite{hk88}) and \object{PSR~J0437-4715} (Bell et al.\ \cite{bbm+95})
argue against the required extreme anisotropies.

{If the above reasoning is correct, we are left with the
possibility of a spin-down luminosity lower than inferred from the
spin period and its derivative.  This could happen if} the spin period
derivative does not reflect actual spin-down, but rather an (apparent)
acceleration, of order $a/c=\dot P/P=4.6\times10^{-17}{\rm\,s^{-1}}$.
Expected contributions to $\dot P$ arise from proper motion,
acceleration in the cluster and differential galactic acceleration
between the cluster and us.  Using estimates from Phinney
(\cite{phi92}), we find that the Galactic acceleration is far too
small, $\sim\!1.6\times10^{-19}{\rm\,s^{-1}}$, while the required
proper motion, $(c\dot P/Pd)^{1/2} \simeq 85{\rm\,mas\,yr^{-1}}$ or
$\sim\!1000{\rm\,km\,s^{-1}}$ at the distance of \object{NGC~6397}, is
outrageously high (and excluded by the HST observations).

The maximum acceleration in a cluster is approximately
$1.1G\Sigma_M/c$ (Phinney \cite{phi92}), where $\Sigma_M$ is the
surface density enclosed within the projected distance of the pulsar
from the core.  With an extinction-corrected V-band surface brightness
within 0\farcm55 of $\Sigma_{V,0}=18.0{\rm\,mag\,arcsec^{-2}}$ (Trager
et al.\ \cite{tkd95}), and a mass to light ratio of about 3 (in solar
units), we find a maximum acceleration of
$\sim\!3\times10^{-18}{\rm\,s^{-1}}$, again far short of that required
(as found by D'Amico et al.\ [\cite{dapm+01}] using a different
method).  We note that the mass to light ratio is uncertain; for
instance, from {\em negative} period derivatives of two pulsars in
\object{NGC~6752}, D'Amico et al.\ (\cite{dapf+02}) infer a mass to
light ratio of $\ga\!10$.  In order for the acceleration to match the
period derivative for \object{PSR~J1740$-$5340}, however, a mass to
light ratio of $\sim\!50$ would be required.  This seems excessive.

\subsection{A third body?}

Having excluded all likely causes, what remains, however improbable?
One possibility is that the binary is passing another star close by or
that it is part of a triple system, and happens to be seen being
accelerated away from us.  This would not be unprecedented:
\object{PSR~B1620$-$26} in \object{M\,4} has a $\sim\!0.3\,M_\odot$
white-dwarf companion in a 191\,d orbit, as well as a
$\sim\!0.007\,M_\odot$ giant planet or brown dwarf like object in a
$\sim\!300{\rm\,yr}$, $e\simeq0.45$ orbit (Ford et al.\
\cite{fjrz00}).  For \object{PSR~J1740$-$5340}, a third object with
mass $M_3$ at separation $a_3=(fGM_3P/\dot Pc)^{1/2} =
200\,(fM_3/0.1M_\odot)^{1/2}{\rm\,AU}$ would be required; here $f =
\sin i_3\sin\phi_3$ with $i_3$ the inclination and $\phi_3$ the
orbital phase relative to the ascending node (a positive acceleration
the third object to be behind the pulsar binary, or $\sin\phi_3>0$).
The corresponding period (or timescale in case of a close encounter)
is $P_3 = 2\pi(a^3/GM)^{1/2} \simeq
400(fM_3/0.1M_\odot)^{3/2}(M/1.7\,M_\odot)^{-1/2}{\rm\,yr}$.  In order
for this to be substantially longer than the current observing time
span of 1.1\,yr, say $P_3>7\,$yr, one requires $M_3>0.007\,M_\odot$.

The separation corresponds to an angular separation $\theta_3 =
(1-f^2)^{1/2}a_3/d = 0\farcs08(f(1-f^2)M_3/0.1M_\odot)^{1/2}$ (note
that $0\leq(f(1-f^2))^{1/2}\leq0.62$).  A more massive third object
might be visible in the {\em HST} images (if not a white dwarf);
however, we found no evidence for a close-by object.  The closest
object is star~B of Ferraro et al.\ (\cite{fpdas01}); with a mass of
about a turn-off mass and a separation of $\sim1\farcs4$, it is too
far away to produce the required acceleration.

\subsection{X-ray luminosity}

An independent measure of the spin-down luminosity can be obtained
from the X-ray luminosity, of $\sim\!8\times10^{30}{\rm\,erg\,s^{-1}}$
(Grindlay et al.\ \cite{ghe+01}).  The fraction of the spin-down
luminosity that is converted into X rays is not understood
theoretically, but one can use other MSPs to make an
empirical calibration.  From the data collected by Possenti et al.\
(\cite{pccm02}; esp.\ Fig.~2a), we find that for the observed X-ray
luminosity, a spin-down luminosity of about
$10^{34.2\pm0.6}{\rm\,erg\,s^{-1}}$ is expected.  The best guess is
about an order of magnitude lower than that inferred from the measured
period and period derivative, and thus supports the suggestion that
the intrinsic period derivative is lower than the measured one.  We
stress, however, that the estimate is very uncertain.  Furthermore,
while the MSP in \object{M~28} follows the relation
defined by MSPs in the field, the sources in \object{47
Tuc} seem somewhat less X-ray bright than expected from that relation
(Grindlay et al.\ \cite{ghec02}).  Indeed, the latter define a
different relation which, if extrapolated, fits the observed
properties of \PSR.

The situation is complicated further by the fact that the companion
might contribute to the observed X-ray flux.  For rapidly rotating,
active stars in and out of binaries, the X-ray luminosity saturates at
$\sim\!0.1$\% of the bolometric luminosity (e.g., Stauffer et al.\
\cite{sta++94}; Patten \& Simon \cite{ps96}; Dempsey et al.\
\cite{dlfs97}).  For the companion of \object{PSR~J1740$-$5340},
$L_{\rm bol}\simeq2.0\,L_\odot$, and hence its X-ray flux could be as
high as $\sim\!8\times10^{30}{\rm\,erg\,s^{-1}}$, i.e., 100\% of the
flux observed.  If the companion does contribute to the observed flux,
this might also explain the observed variability (Grindlay et al.\
\cite{ghe+01}).  Interestingly, comparing its relatively hard X-ray
colour with BY Dra systems (which contain an active, rapidly rotating
star) and other MSPs in \object{NGC~6397} and \object{47 Tuc}
(Grindlay et al.\ \cite{ghem01}, \cite{ghe+01}), it seems more
consistent with the former.  Indeed, as mentioned in
Sect.~\ref{intro}, Taylor et al.\ (\cite{tgec01}) had, when the
identification with \PSR\ was not yet known, classified the companion
as a BY~Dra system.

\subsection{Ramifications}

D'Amico et al.\ (\cite{dapm+01}) suggested that, as in the black-widow
pulsars, the radio eclipses were due to mass loss driven by pulsar
irradiation.  Given the observed lack of heating of the companion,
this seems rather implausible.  Instead, we think that an intrinsic
wind of the companion is responsible, as has been suggested for the
very unenergetic eclipsing pulsar \object{PSR~B1718$-$19} in NGC~6342
(Wijers \& Paczynski \cite{wp93}).  Such a wind can be sufficiently
strong because of the companion's rapid (co)rotation.

If the lack of heating is indeed due to an intrinsic spin-down rate
lower than that observed, this affects the inferred values of the
magnetic field strength and characteristic age of the pulsar.  From
our observations, we infer a spin-down luminosity a factor $>\!10$
smaller than inferred from the period derivative, implying that the
intrinsic period derivative $\dot P_{\rm int}> \dot P_{\rm obs}/10
\simeq 2\times10^{-20}$.  Hence, the inferred magnetic field strength
would become $<\!3\times10^8{\rm\,G}$ and the characteristic age
$>\!3\times10^9{\rm\,yr}$.

\section{Current and evolutionary state}

Based on our analysis of {\em HST} photometry of \PSR, we found that
the companion has radius $R_{\rm comp}=1.60\pm 0.17\,R_{\odot}$ and
temperature of $T_{\rm eff}=5410\pm 50$~K, implying a luminosity
$L_{\rm bol}=2.0\pm0.4\,L_\odot$ {(for the Gratton et al.\
distance, the radius and luminosity would be $1.52\pm0.08\,R_\odot$
and $1.8\pm0.2\,L_\odot$, respectively)}.  From the ellipsoidal
variations, we infer that the system's inclination is $>\!48^\circ$.
For a pulsar mass in the range $1.2<M_{\rm MSP}<2.4\,M_{\odot}$, this
implies a companion mass in the range $0.14<M_{\rm
comp}<0.38\,M_{\odot}$.  In this range, the companion is just short of
filling its Roche lobe (at just over the 2-$\sigma$ level, however, it
may be filling its Roche lobe completely).

From our UVES spectrum, we find that the companion has low
metallicity, as expected for membership of \NGC, that its radial
velocity is consistent with membership for a mass ratio in the range
$5<Q<7$, and that it rotates rapidly, $v_{\rm rot}\sin
i=52\pm4{\rm\,km\,s^{-1}}$, more or less consistent with it rotating
synchronously with the orbit.

The largest surprise was that the light curve showed no evidence for
heating.  We discussed this at length and by an admittedly long chain
of reasoning concluded that most likely the measured period derivative
was dominated not by intrinsic spin-down, but by acceleration from a
third object orbiting or passing by the system.

The picture that emerges is one of a system somewhat like an RS~CVn
binary, in which the co-rotating companion of the pulsar is an active
star with a relatively strong but variable wind, which disperses and
absorbs the pulsar signal for extended periods, in particular around
superior conjunction of the pulsar.  With this, we suggested an answer
to the third of the three puzzles mentioned in the introduction, viz.,
the origin of the material causing the eclipses.  We will now discuss
the remaining two.

\subsection{Location in the cluster}

The location of the system is puzzling because it is far out of the
core, at $0\farcm55$ or eleven core radii, even though the total mass
of $\ga\!1.6\,M_\odot$ is well above the turn-off mass of
$0.8\,M_\odot$.  At the current position, the mass segregation time is
rather short, between 1 and 10\,Myr depending on how close to the core
the system is brought in its current orbit (Binney \& Tremaine
\cite{bt87}; Spitzer \cite{spi87}; we used a system mass of
$1.7\,M_\odot$, current radius of 0.4\,pc, enclosed mass of
$2000\,M_\odot$, and total number of objects in the cluster of
$3\times10^5$).  {This suggests that the system was kicked out
of the core, presumably by a close encounter.  Probably, also, this
interaction happened not too long ago, since it seems unlikely that
the system was kicked out very far, and happens to be in the last
stages of sinking back to the core at the present time.  On the other
hand, if the system really is young, one may wonder why not many more
similar systems are known.  Independent of the timescale,} if there is
indeed a third component in the current system, the interaction must
have been between (at least) four components, the remaining object(s)
to have left the core in the direction opposite to that of the pulsar
triple system.

\subsection{Properties of the companion}

The companion of \object{PSR~J1740$-$5350} clearly is an unusual star:
given its low mass, it should be on the main sequence, but it is far
too large and bright for that.  Might the star have been more massive
originally?  If so, did it lose its mass gradually, due to binary
evolution, or impulsively, in an encounter?  Alternatively, could the
star have gotten bloated?  We discuss these options briefly in turn.

\subsubsection{Binary evolution}

If the star originally was more massive, it could have lost its mass
due to mass transfer in `regular' binary evolution.  This has been
suggested by Burderi et al.\ (\cite{bdab02}) and Ergma \& Sarna
(\cite{es02}).  Due to the mass transfer, the neutron star is expected
to be spun up to millisecond periods.  As the suggested solutions
stand, there are some small problems: Ergma \& Sarna include heating
by the pulsar, for which we see no evidence, and Burderi et al.\ end
up with a slightly too massive companion, of $0.45\,M_\odot$.  We have
little doubt, however, that suitable tuning could remedy this.
{Furthermore, in both models, the mass-transfer rate is only a
few $10^{-10}\,M_\odot{\rm\,yr^{-1}}$, which implies a mass-transfer
stage of several Gyr following its ejection from the core.  Thus, one
has to appeal to a coincidence that we happen to observe it in its
last few Myr of sinking back.}

A more pressing problem seems to be that in both models the companion
is expected to continue to fill its Roche lobe.  This is only
marginally consistent with the fit to the light curve.  Furthermore,
it requires the assumption that, somehow, at some point the MSP
switches on and manages to blow away all the matter thrown at it
(which is a substantial amount in the model of Burderi et al.\
[\cite{bdab02}]).  This would seem to require an energetic pulsar;
however, we see no evidence for this radiation hitting the companion.

\subsubsection{A stripped turn-off star}

Could the encounter that kicked the system out of the core also have
caused the companion to reach its present state?  A possibility is
that during a multiple-object interaction, the neutron star passed a
star very closely, within about three stellar radii, which caused the
star to be partially disrupted.  Such a scenario was suggested by
Zwitter (\cite{zwi93}) for the eclipsing pulsar
\object{PSR~B1718$-$19}.  In order for the disruption not to be
complete, the star would have had to have a well-developed core, i.e.,
be a sub-giant or giant.  During the interaction, some mass may been
accreted by the neutron star.  This could explain the 1 second period
of \object{PSR~B1718$-$19}, but would likely be insufficient to spin
up a neutron star to the 3.65 ms period observed for \PSR.  If so, the
neutron must have been a MSP already before the interaction.

It is not quite clear how a star reacts after rapidly losing more than
half its mass, likely followed by a phase of strong tidal dissipation.
Still, given that the luminosity for (sub-)giants depends mostly on
the mass of the core, which would not change, one might suppose ending
up with a star with roughly the same luminosity as it had before.  For
\object{PSR~J1740$-$5340}, the luminosity of the companion is indeed
comparable to the turn-off luminosity.  Unfortunately, no prediction
seems yet possible for the temperature or radius.

\subsubsection{A bloated main-sequence star}

For \object{PSR~B1718$-$19}, the scenario invoking a stripped turn-off
star cannot be correct, as the companion luminosity is far too low
(Van Kerkwijk et al.\ \cite{vkkk+00}).  Instead, the observations are
consistent with another scenario, suggested by Wijers \& Paczynski
(\cite{wp93}), viz., that the companion interacted with a neutron star
in a less destructive manner, but was bloated during the subsequent
circularisation phase.  Could the same be true for
\object{PSR~J1740$-$5340}?  We see two possible problems.  First,
since the orbit has been circularised, there are no energy sources any
more in the companion except for nuclear processes.  Hence, the
companion should be shrinking on its thermal timescale, which is
rather short, $GM^2/RL=1.0\times10^6{\rm\,yr}$.

The second problem is that, in contrast to what is the case for
\object{PSR~B1718$-$19} (Verbunt \cite{ver94}), the difference in
total energy between the original eccentric orbit and the current
circular one, $|E_{\rm circ}|-|E_{\rm ecc}|<|E_{\rm circ}| = GM_{\rm
MSP}M_{\rm comp}/2a = 1.2\times10^{47}{\rm\,erg}$, is substantially
smaller than the binding energy of
$\frac{6}{7}GM^2/R=1.0\times10^{48}{\rm\,erg}$ for a completely
convective, $0.3\,M_\odot$, $0.3\,R_\odot$ main-sequence star.  This
makes it doubtful that the star could be bloated sufficiently.
Bloating of just the outer layers would not help: at the present
luminosity, these would shrink back very rapidly.

\subsection{Sub-subgiants and red stragglers}

The combination of luminosity and temperature shown by the companion
of \object{PSR~J1740$-$5350} is unusual, but not unique.  Edmonds et
al.\ (\cite{egc+02}) pointed out the similarity with a number of
objects found in \object{47~Tuc} by Albrow et al.\ (\cite{agb+01}).
Albrow et al.\ dubbed these `red stragglers,' and noted that their
properties are similar to two so-called `sub-subgiants' in the old
open cluster \object{M\,67}, which are discussed in detail by Mathieu
et al.\ (\cite{mvdbt+02}).  All these objects are found well to the
red of the main sequence, and about half a magnitude below the bottom
of the red-giant branch.  For all, the implied luminosities are
comparable to those of turn-off stars, like for
\object{PSR~J1740$-$5340}.

Mathieu et al.\ (\cite{mvdbt+02}) note that the two \object{M\,67}
sub-subgiants most likely are in or close to thermal equilibrium,
since given the short thermal timescales it would otherwise be
unlikely to find two in a (relatively) sparse cluster like
\object{M\,67}.  Albrow et al.\ (\cite{agb+01}) suggest that for all
sources the peculiar properties are due to mass transfer driven by
evolution.  This can indeed produce such stars (as shown also by the
models for \object{PSR~J1740$-$5340} discussed above).  For both
objects in \object{M\,67}, however, this is not possible: one is in an
eccentric orbit, and the other is underfilling its Roche lobe (Mathieu
et al.\ \cite{mvdbt+02}).

Could these sub-subgiants result from a close encounter?  Since the
luminosities are so close to those of turn-off stars, it seems
unlikely they are tidally bloated stars (as also suggested by the
thermal timescale argument).  In our list of scenarios for
\object{PSR~J1740$-$5340} (which may of course be incomplete), this
leaves only the possibility of stripped (sub-)giant.  As mentioned, it
is at present not clear whether such a stripping would lead to a star
with the observed temperature, and to what extent the properties would
depend on, e.g., the actual amount of mass loss.  

The observed positions in the colour-magnitude diagram, close to the
extension of the giant branch, might be taken to suggest the star has
turned into a giant, but of lower mass than is possible by normal
evolution.  If so, since the lowest possible core masses for
(sub-)giants are similar even among different (old) clusters, the
similarity in their (starting) luminosities follows naturally.  Also,
since the further evolution would be on the slow nuclear timescale,
finding relatively many would be possible.  Finally, if this
explanation holds, `red stragglers' would be an appropriate name: just
like blue stragglers are not strange in their properties as such, but
rather in their apparent youth (relative to what should be their
contemporaries), the red stragglers would be strange in their apparent
old age.

\section{Future work}

We have shown that it is possible to obtain reasonably good echelle
spectra of the companion star in \PSR.  The next obvious step is to
obtain several more spectra and measure the full radial velocity curve
of the companion star.  It should not be unduly difficult to obtain
an accuracy of $K_{\rm comp}$ well below $1{\rm\,km\,s^{-1}}$, which
in turn will yield a mass ratio accurate at well below the 1\% level.
Apart from the double neutron star binaries, the mass ratio of \PSR\
would be the most accurately known mass ratio of a binary with a
compact object.

Given that the mass ratio $Q$ can in principle be known extremely
accurately, the only remaining quantity needed to obtain good
component masses is the inclination $i$.  In this case we are at a
slight disadvantage compared to, e.g., studies of X-ray binaries,
since we cannot assume the companion fills its Roche lobe for the
purposes of ellipsoidal modelling.  As a result there is a large range
of inclinations where the fits to the light curves are good
(Fig.~\ref{plotsincont}).  We must use other constraints to narrow
down this range, namely the radius of the companion from the {\em HST}
photometry and the rotational velocity of the companion.  Improvement
of the former requires a better distance, {which may already
have become available with the detailed work of Gratton et al.\
(\cite{gcb+02});} for a very precise distance, one will have to wait
for dedicated astrometric space missions.  The precision of the
measurement of the rotational velocity of the companion star, however,
will improve with additional echelle spectra.  From
Fig.~\ref{plotsincont}f, one sees that very high accuracy is required:
even given a precisely known value of $Q$, an uncertainty in $v_{\rm
rot}\sin i$ of $1{\rm\,km\,s^{-1}}$ corresponds to a final uncertainty
in the mass of $\sim\!0.2\,M_\odot$.

The allowed inclination range can also be reduced by using light
curves with better phase coverage and higher statistical quality.
Observations over several binary orbits should be obtained in order to
{verify that the light curves are indeed ellipsoidal}, check
that they are stable, and look for the presence of short-period
variations such as those caused by star spots (if present, these could
be used to verify the assumption of co-rotation).  \NGC\ has been well
studied, and a large number of suitable observations no doubt exist in
personal archives (e.g.\ Rubenstein \& Bailyn \cite{rb96}; Kaluzny
\cite{kal97}), although likely advanced techniques such as image
subtraction will be needed to recover good light curves.

It should be relatively easy to obtain an average echelle spectrum of
the companion star with a signal-to-noise ratio on the order of~100.
This would allow one to do a detailed abundance analysis, and search
for clues to the evolution of this system; e.g.\ a captured and
perturbed low-mass main sequence companion will have a different
composition than an initially higher-mass star that has lost a
substantial amount of mass (e.g., Ergma \& Sarna \cite{es02}).

Finally, further radio timing, as well as detailed inspection of a
good set of spectra, would allow one to search for evidence for the
third star that we suggested was present.  Higher signal-to-noise and
time resolved X-ray observations might clarify whether the X-ray
emission arises from the pulsar or from the companion.

{While revising this article, we became aware of a preprint of
Kaluzny et al.\ (\cite{krt02}) reporting ground-based photometry and
low-resolution spectroscopy of the companion of \PSR.  They also find
that the lightcurves, which have good phase coverage, are well
described by ellipsoidal modulation.  In detail, however, the results
show small but significant differences from ours, mostly because their
2002 lightcurves turns out to have slightly smaller amplitude than our
1995 {\em HST} lightcurves.  The main conclusions of our work,
however, remain unchanged, in particular that there is no evidence for
irradiation.  Kaluzny et al.\ discuss the change in lightcurve
amplitude and mention that it could reflect a change in inclination
due to a third body, which we invoked above to explain the lack of
irradiation.  As said, this hypothesis can be tested with further
timing.  A more mundane and therefore perhaps better solution, also
suggested by Kaluzny et al., is the presence of star spots.  Finally,
Kaluzny et al.\ measure radial velocities and derive a value for the
radial-velocity amplitude, of $137.2\pm2.4{\rm\,km\,2^{-1}}$.  This
would imply $Q=5.15$.  This value is consistent with the limits
derived above, but significantly lower than our preliminary result
based on further UVES observations, possibly because of blending with
other stars in the bad-seeing conditions Kaluzny et al.\ were faced
with.}

\begin{acknowledgements}
We thank the staff at Paranal Observatory for performing the observations
in a timely fashion, and the staff at ESO Garching for delivering the
data to us in an expedited manner.  This research made extensive use
of ADS and SIMBAD.  MHvK is supported by a fellowship of the Royal
Netherlands Academy of Science.
\end{acknowledgements}

\end{document}